\documentclass[12pt]{article}
\usepackage[xdvi]{graphicx} 
\usepackage{epsfig}
\usepackage{amsfonts,amssymb,amsmath}

\newcommand{\al}{\ensuremath{\alpha}}

\newcommand{\ep}{\ensuremath{\epsilon}}

\newcommand{\ka}{\ensuremath{\kappa}}
\newcommand{\la}{\ensuremath{\lambda}}
\newcommand{\La}{\ensuremath{\Lambda}}

\renewcommand{\d}{\ensuremath{{\rm d}}}

\newcommand{\Del}{\ensuremath{\nabla}}

\newcommand{\be}{\begin{equation}}
\newcommand{\ee}{\end{equation}}

\newcommand{\ba}{\begin{eqnarray}}
\newcommand{\ea}{\end{eqnarray}}

\begin{document}

\rightline{hep-th/0406157}
\rightline{OUTP-04/13P}
\vskip 1cm 

%% 			Title here 
%%
\begin{center}
{\Large \bf Cosmic acceleration from asymmetric branes}
\end{center}
\vskip 1cm
  
\renewcommand{\thefootnote}{\fnsymbol{footnote}}

\centerline{\bf Antonio Padilla\footnote{a.padilla1@physics.ox.ac.uk}}
\vskip .5cm

\centerline{\it Theoretical Physics, Department of Physics}
\centerline{\it University of Oxford, 1 Keble Road, Oxford, OX1 3NP,  UK}

\setcounter{footnote}{0} \renewcommand{\thefootnote}{\arabic{footnote}}
 
%%			Text starts here 
%% 

\begin{abstract}
We consider a single 3-brane sitting in between two different five
dimensional spacetimes. On each side of the brane, the bulk is a
solution to Gauss-Bonnet gravity, although the bare cosmological constant, fundamental Planck
scale, and Gauss-Bonnet coupling can differ. This asymmetry leads to
weighted junction conditions across the brane and interesting brane
cosmology.  We focus on two special cases: a generalized
Randall-Sundrum model without any Gauss-Bonnet terms, and a stringy
model, without any bare cosmological constants, and positive
Gauss-Bonnet coupling. Even though we assume there is no vacuum energy
on the brane, we find late time de Sitter cosmologies can
occur. Remarkably, in certain parameter regions, this acceleration is
preceded by a period of matter/radiation domination, with $H^2 \propto
\rho$, all the way back to nucleosynthesis.
\end{abstract}

\newpage

\section{Introduction}
Recent observations suggest that our universe is
accelerating~\cite{Perlmutter:accn,Riess:accn}. In the standard cosmology, such an acceleration
cannot be driven by ordinary matter and radiation. The simplest
explanation is to imagine that there is some form of positive vacuum
energy/cosmological constant. Although some progress has been made
recently~\cite{Kachru:dS, Kachru:inflation, Bala:accn}, it is notoriously difficult to produce a
positive cosmological constant from compactifications of
string/M-theory. Given that string theory is currently our best candidate for a
quantum theory of gravity, this is a big worry. It is natural,
therefore, to seek other explanations.

The standard technique is to modify Einstein gravity is some
particular way. In quintessence theories we add a scalar field to
obtain the desired acceleration~\cite{Zlatev:quin}. However, none of these theories
appear to be related to a more fundamental theory of quantum
gravity. Another solution is to consider theories that exhibit
infra-red modifications of gravity~\cite{Dvali:DGP, Kogan:multi5, Papazoglou:thesis, Gregory:GRS,Arkani-Hamed:ghost1,Arkani-Hamed:ghost2}. Generically, these also
lead to cosmic acceleration at late times~\cite{Deffayet:cosmo,Lue:accn,Dvali:power,Damour:accn, Holdom:accn,Sahni:bwde}. Perhaps the most
celebrated of these theories are the DGP model~\cite{Dvali:DGP}, 
multigravity~\cite{Papazoglou:thesis}, and more recently the idea of a ghost condensate~\cite{Arkani-Hamed:ghost1,Arkani-Hamed:ghost2}. The DGP model is a braneworld model where there
is a large amount of curvature induced on the brane. It has been argued that this model suffers
from a strong coupling problem~\cite{Luty:stong,Rubakov:strong}, although the jury is still out
in some respect~\cite{Dvali:IR}. For their part, multigravity  models are
often plagued by ghosts~\cite{Papazoglou:thesis}. One can construct a DGP-like model that
exhibits {\it bi}gravity and is free from ghosts~\cite{Padilla:bigravity}, but the strong coupling
problem still looms large. The ghost condensate, on the other hand,
is an exotic form of matter whose dispersion relation has the form
$\omega^2 \propto k^4$, owing to Lorentz symmetry breaking. This model
is very interesting, and can successfully describe cosmic
acceleration. However, it is a low energy effective theory, and as yet
we have
no insight into the UV completion beyond the Lorentz
symmetry breaking scale.

In this paper, we will suggest an alternative to each  of the
above. We will consider a single 3-brane that acts as a
domain wall between two different five dimensional spacetimes. These
bulk spacetimes will, in general, be solutions to Gauss-Bonnet
gravity. This is the combination of the Einstein-Hilbert action and
the Gauss-Bonnet term,
\be \label{eq:bulkaction}
S = M^3\int_\mathcal{M} \d^5 x
\sqrt{-g}\lbrace R - 2\La + \al \mathcal{L}_{\mathrm{GB}}\rbrace,
\ee
where
\be
\mathcal{L}_{\mathrm{GB}} = R^2 - 4 R_{ab}R^{ab} + R_{abcd}R^{abcd}.
\ee
In 4 dimensions, a linear combination of the Einstein tensor and the
metric is the most general combination of tensors satisfying the
following conditions
\begin{itemize}
\item it is symmetric.
\item it depends only on the metric and its first two derivatives.
\item it has vanishing divergence.
\item it is linear in the second derivatives of the metric\footnote{In
4 dimensions this condition is actually implied by the other three.}.
\end{itemize}
If we go to 5 or 6 dimensions, it turns out
that these conditions are satisfied by a linear combination of the
metric, the Einstein tensor, and the {\it Lovelock
tensor}~\cite{Lovelock:einstein,Lanczos}. The Lovelock tensor arises
from the variation of the Gauss-Bonnet term in the above
action~(\ref{eq:bulkaction}). In this sense, Gauss-Bonnet gravity is the natural
generalisation of Einstein gravity to higher dimensions.

String theory provides us with an even more compelling reason to study
Gauss-Bonnet gravity, especially in a braneworld context. In the Regge
slope ($\al'$ ) expansion of the heterotic string
action, curvature
squared terms appear as the leading order correction to
Einstein gravity~\cite{Candelas:vac, Green:Dinst}.  Furthermore, for this theory
of gravity to be ghost-free, the curvature squared terms must appear
in the Gauss-Bonnet combination~\cite{Zwiebach:gb, Zumino:theories,
Gross:heterotic}.

Brane cosmologies with and without a Gauss-Bonnet correction have been
extensively studied (see, for example~\cite{Randall:RS1, Randall:RS2, Binetruy:cosmo,Bowcock:cosmo, Maartens:bwgravity,Langlois:intro, Brax:review,
Padilla:thesis,Charmousis:GBcosmo,Charmousis:scalar, Kofinas:cosmo, Papantonopoulos:cosmo,Padilla:gbhol, Dehghani:accn}). Generically, to obtain a de Sitter phase
of cosmological expansion one needs to introduce a positive vacuum
energy in the bulk or on
the brane~\cite{Padilla:thesis,Karch:locally,Gen:dS,Padilla:CFT}. The alternative is to include some induced brane curvature~\cite{Deffayet:cosmo,Lue:accn,Dvali:power, Kofinas:cosmo}. It is our desire to avoid doing either of these for the reasons
outlined above. We should note, however, that in~\cite{Kofinas:cosmo}, the
authors also manage to avoid an initial singularity through the
combined effect of the induced curvature and the Gauss-Bonnet
bulk. Although this will not happen in any of our models, we will
suggest an alternative means of doing this in
section~\ref{Discussion}. In~\cite{Dehghani:accn}, cosmic acceleration is achieved by
considering negative Gauss-Bonnet coupling ($\al<0$), but this is not
well motivated by string theory, and has problems with
stability~\cite{Boulware:string}. 

The key feature in our model is the asymmetry across the brane. This
means that the parameters in our theory can differ on either side of
the brane. This has previously been applied to the bulk energy
momemtum and the bulk Weyl
tensor~\cite{Bowcock:cosmo,Charmousis:GBcosmo,Charmousis:scalar,Davis:Z2,Carter:reflection,Battye:jc,Battye:asymm,Carter:asymm,Melfo:thick,Gergely:gen,Castillo:local}. Here we will also apply it to the gravitational
couplings, as in~\cite{Stoica:cosmo}. There are at least two ways in
which this asymmetry might arise. Firstly, suppose we have some sort of wine bottle shaped
compactification down to 5 dimensions. In the effective theory, the
Planck scale at the fat end  of the bottle will be less than that at
the thin end. Secondly, in~\cite{Padilla:nested, Padilla:instantons}
we showed how to construct a domain wall living entirely on the
brane. In some cases~\cite{Padilla:instantons}, the Planck scale on
the brane differed on either side of the domain wall.

The jump in the gravitational couplings leads to weighted junction
conditions across the brane. In the absence of any vacuum energy on
the brane, it is this fact that enables us to find
late time de Sitter solutions, with $H^2 \sim (constant)^2$.  However, this alone is not enough to describe a realistic
cosmology. The accelerated expansion must be preceded by an era of
matter and radiation domination, with $H^2 \propto \rho$, all the way
back to nucleosynthesis. We will
show that this can also be achieved, at least in certain parameter
regions. This is perhaps the most remarkable aspect of our work.

The rest of this paper is organised as follows: in
section~\ref{Equations} we will introduce our asymmetric brane action,
and show how to construct a homogeneous and isotropic braneworld. In
section~\ref{Einstein}, we will study a generalized Randall-Sundrum
model, for which the bare
cosmological constants are negative ($\La<0$) and there are no
Gauss-Bonnet couplings ($\al=0$). In section~\ref{stringy} we will
consider  a more stringy example. We will assume that there are no
bare cosmological constants ($\La=0$), so that the bulk action  contains only the
Ricci scalar and  Gauss-Bonnet terms. This is what we
would expect from the slope expansion of heterotic string theory.
Section~\ref{Discussion} contains some concluding remarks.

\section{Equations of motion} \label{Equations}

Consider two 5 dimensional spacetimes, $\mathcal{M}_1$ and
$\mathcal{M}_2$, separated by a domain wall. The domain wall is
a 3-brane corresponding to our universe. In general, $\mathcal{M}_i$
is a solution to Gauss-Bonnet gravity
with a (bare) cosmological constant, $\Lambda_i$,
and Gauss-Bonnet coupling $\alpha_i$. The 5-dimensional
Planck scale in this region will be given by $M_i$. We will not
assume that there is $\mathbb{Z}_2$ symmetry across the brane, so that
the fundamental parameters of our theory can differ on either side of
the brane. This scenario is described
by the following action,
\be \label{action}
S= S_\textrm{grav}+S_\textrm{brane},
\ee
where
\ba
S_\textrm{grav} &=& \sum_{i=1, 2} M_i^3\int_{\mathcal{M}_i} \d^5 x
  \sqrt{-g}\lbrace R - 2\La_i + \al_i \mathcal{L}_{\mathrm{GB}}\rbrace
  + \int_{\partial \mathcal{M}_i}
  \textrm{boundary terms} \\
S_\textrm{brane} &=& \int_{\textrm{brane}} d^4 x
  \sqrt{-h}\, \mathcal{L}_\textrm{brane}.
\ea
The boundary integrals in $S_\textrm{grav}$ are required for a well
defined action principle~\cite{Myers:ghterm} (see
also~\cite{Davis:israel}). We denote the bulk  metric and the brane
metric by $g_{ab}$ and $h_{ab}$
respectively. $\mathcal{L}_\textrm{brane}$ describes the matter
content on the brane. We will assume that this is made up of ordinary
matter and radiation, so that there is no vacuum energy. In other
words, there is no brane tension.

\subsection{The bulk}

In $\mathcal{M}_i$,  the bulk equations of motion are given
by
\be
R_{ab} - \frac{1}{2}R g_{ab} = -\La_i g_{ab} + \al_i
  \Big\{ \frac{1}{2}\mathcal{L}_{\mathrm{GB}}\,g_{ab}
    -2R R_{ab}+4R_{ac}{R_b}^c+4R_{acbd}R^{cd}-2R_{acde}{R_b}^{cde}
  \Big\}
\ee
For the time being, let us drop the index $i$, as the following
analysis will apply on both sides of the brane. We will put it back in
when necessary.

If we demand that the bulk contains 3-dimensional spatial sections of
constant curvature, we find the following
solutions~\cite{Charmousis:GBcosmo, Boulware:string, Cai:GBBH},
\be \label{eq:gbbh}
\d s^2 =-h(a) \d t^2 +\frac{\d a^2}{h(a)}
  + a^2 \d {\bf x}_\ka^2,
\ee
where
\be \label{metric}
h(a)  =  \ka + \frac{a^2}{4 \al}\left(1 \pm \xi(a) \right)
  \quad \textrm{with} \quad \xi(a) =  \sqrt{1 +\frac{4\al\La}{3}  
  + \frac{8 \al\mu}{a^{4}}}.
\ee
For $\ka=1, 0, -1$, $\d {\bf x}_\ka^2$ is the metric on a unit 3-sphere,
plane, and hyperboloid respectively. $\mu$ is a constant of
integration. Other solutions do exist for special values of $\ka$, $\La$, and $\al$~\cite{Charmousis:GBcosmo}, but we
will not consider them here.

For the metric to be real for all $0 \leqslant a < \infty$, we
clearly require that
\ba
1 +\frac{4\al\La}{3} &\geqslant 0& \label{cond0}\\
\al \mu &\geqslant 0&  \label{cond2}
\ea
Furthermore, the mass of the spacetime is given by~\cite{Padilla:hamiltonian, Crisostomo:BHscan, Deser:quadenergy,
Deser:HDenergy}
\be
m=3M^3V\mu,
\ee
where $V$ is the volume of the homogeneous sections. In order to
avoid a classical instability, we must have $\mu
\geqslant 0$. This is in some sense counter-intuitive as the (+) branch
of~(\ref{metric}) then asymptotes to a Schwarzschild metric with {\it
negative} mass, if one uses the standard ADM formula for Einstein
gravity~\cite{Brown:energy, Hawking:ham}. 

From equation~(\ref{cond2}), we see that for $\mu > 0$, we must have
$\al \geqslant 0$, which is consistent with string theory. On a less
positive note, we also find that the metric has a singularity at
$a=0$. For the (-) branch, this singularity is covered by an event
horizon. This is not the case for the (+) branch. To shield this naked
singularity we must cut the spacetime off at some small value of $a$.
This can be done by introducing a second brane at, say,  $a
\sim M^{-1}$~\cite{Brax:naked}.

\subsection{The brane}

In order to construct a brane, we glue a solution in $\mathcal{M}_1$
to a solution in  $\mathcal{M}_2$, with the brane forming the common
boundary. Let us describe this in more detail. In
$\mathcal{M}_i$,  the boundary, $\partial \mathcal{M}_i$, is given by the  section $(t_i(\tau), a_i(\tau), {\bf
x}^{\mu})$ of the bulk metric. The parameter
$\tau$ is the proper time of an observer comoving with the boundary, so that
\be \label{unit}
-h_i(a_i)\dot{t}^2 + \frac{\dot{a}^2}{h_i(a_i)} = -1,
\ee
where overdot corresponds to differentiation with respect to
$\tau$. The {\it outward} pointing unit normal to $\partial \mathcal{M}_i$
is now given by
\be
n_a=\theta_i( -\dot a_i(\tau), \dot t_i(\tau), {\bf 0})
\ee
where $\theta_i= \pm 1$. For $\theta_i=-1$, $\mathcal{M}_i$
corresponds to  $a_i(\tau) < a< \infty$, whereas for  $\theta_i=
1$, $\mathcal{M}_i$ corresponds to  $0 \leqslant a < a_i(\tau)$.

The induced metric on $\partial \mathcal{M}_i$ is that of a FRW universe,
\be \label{eq:frw}
\d s^2 = -\d \tau^2 + a_i(\tau)^2 \d {\bf x}_\ka^2,
\ee
Since the brane coincides with both boundaries, the metric on the
brane is only well defined when $a_1(\tau)=a_2(\tau)=a(\tau)$. The
Hubble parameter on the brane is now defined 
by $H=\dot a/a$.

The dynamics of the brane are determined by the junction conditions for a braneworld in
Gauss-Bonnet gravity~\cite{Davis:israel, Gravanis:israel}. This comes
from varying the action (\ref{action}) with respect to the brane
metric. Given a quantity $Z_i$ defined in $\mathcal{M}_i$, we shall
henceforth write $\langle Z \rangle=(Z_1+Z_2)/2$, for the average across the brane, and $\Delta
Z= Z_1-Z_2$, for the difference. The brane equations of motion are
given by~\cite{Davis:israel, Gravanis:israel}
\be \label{eq:junction}
2\langle X_{ab} \rangle =T_{ab}-\frac{1}{3}T h_{ab}
\ee
where, suppressing the index $i$,
\be
X_{ab} =2M^3\left[K_{ab}+2 \al \left(Q_{ab}-\frac{2}{9}Q h_{ab}
\right) \right].
\ee
Here, 
\ba
Q_{ab} & = & 2K K_{ac}{K_b}^c -2K_{ac}K^{cd}K_{db} +
                K_{ab}(K_{cd}K^{cd}-K^2) \nonumber \\
& & \quad + 2K \mathcal{R}_{ab} + \mathcal{R} K_{ab}
  - 2 K^{cd}\mathcal{R}_{cadb} - 4\mathcal{R}_{ac}{K_b}^c
\ea
and $K_{ab} = h_a^c h_b^d \Del_{\left(c\right. }
n_{\left. d \right)}$ is the   extrinsic  curvature of the
brane in $\mathcal{M}$. $\mathcal{R}_{abcd}$ is the Riemann tensor on
the brane, constructed from the induced metric $h_{ab}$.

The energy-momentum tensor on the brane is given by
\be
T_{ab}=-\frac{2}{\sqrt{-h}}\frac{\delta S_\textrm{brane}}{\delta
  h^{ab}}.
\ee
Since the brane is homogeneous and isotropic
\be
T_{ab} = (\rho+ p)\tau_a \tau_b +
p h_{ab},
\ee
where $\rho$ is the energy density, $p$ is the pressure,  and
$\tau^a$ is the 
velocity of a comoving observer.  Note that in $\mathcal{M}_i$, $\tau^a=(\dot
t_i(\tau), \dot a(\tau), {\bf 0})$, and recall that the unit normal to
$\partial \mathcal{M}_i$ is $n_a =\theta_i
(- \dot{a}(\tau), \dot{t_i}(\tau),{\bf 0})$. We now evaluate the
spatial components of~(\ref{eq:junction}) to give
\be \label{eq:dott}
2\Big\langle \theta M^3\frac{h\dot t}{a}\left[1
-\frac{4}{3}\al\left(\frac{h\dot t}{a}\right)^2 +4 \al
\left(H^2+\frac{\ka}{a^2} \right)\right] \Big\rangle
=\frac{\rho}{6}.
\ee
Making use of equation (\ref{unit}), we can simplify this
expression to give
\be \label{eom1}
2\langle \theta F(H^2) \rangle=\frac{\rho}{6}
\ee
where
\be
F(H^2)=M^3\sqrt{\frac{h}{a^2}+H^2}\left[1
+\frac{8}{3}\al H^2+\frac{4}{3} \al\left(\frac{3\ka-h}{a^2}\right) \right]
\ee
Equation (\ref{eom1}) suggests that the brane dynamics will depend
crucially on the relative signs of $\theta_1$ and $\theta_2$. We shall
therefore consider each case separately. For
$\theta_1=\theta_2=\theta$, we have
\be \label{G+}
\theta \frac{\rho}{6}=G_+(H^2) \equiv 2\langle F(H^2) \rangle
\ee
whereas for $\theta_1=-\theta_2=\theta$, we have 
\be \label{G-}
\theta \frac{\rho}{6}=G_-(H^2) \equiv \Delta F(H^2)  
\ee
These equations are very complicated. However, we can, in principle,
analyse their behaviour, particularly at late times. Recall that we
are assuming that there is {\it no} vacuum energy on the brane. We
might naively expect this to prohibit cosmic acceleration at late
times. In the examples that follow, we will show that this expectation
is wrong, and that we {\it can} get $H^2 \sim (constant)^2$, for large
$a$.  Furthermore, we will show that, in certain parameter regions, this is preceded by an era of
matter and radiation domination, with $H^2 \propto \rho$ as far back
as nucleosynthesis.

The examples we will consider have been chosen both for interest and
simplicity. These are the generalized Randall-Sundrum model and the
stringy model described in the introduction.

\section{The generalized Randall-Sundrum model} \label{Einstein}
In this section we will consider the generalized RS model, for which
\be
\al_i=0 \qquad \La_i=-\frac{6}{l_i^2} 
\ee
If we assume that $M_1 \neq M_2$, we can, without loss of generality,
take $M_1>M_2>0$. For the ``symmetric'' scenario
($\theta_1=\theta_2$), this corresponds to the model discussed
in~\cite{Stoica:cosmo}.

Only the (-) branch of (\ref{metric}) is 
well-defined. It is reduced to the AdS-Schwarzschild metric
\be
h(a)=\ka+\frac{a^2}{l^2}-\frac{\mu}{a^2}
\ee
The equations of motion are given by (\ref{G+}) and (\ref{G-}) with
\be
F(H^2)=M^3\sqrt{\frac{1}{l^2}-\frac{\mu}{a^4}+\frac{\ka}{a^2}+H^2}
\ee
We shall begin by looking for late time de-Sitter solutions, in order
to describe the current cosmic acceleration. We will assume that $a
\to \infty$ at late times. In this limit,
\be
F(H^2) \to M^3\sqrt{\frac{1}{l^2}+H^2}, \qquad \rho \to 0
\ee
where we have used the fact that there is no vacuum energy contained
in $\rho$. Let us start with the ``symmetric'' equation of motion
(\ref{G+}). At late times, it reads
\be \label{latesym}
0=G_+(H^2)\equiv 2\langle F(H^2) \rangle
\ee
It is easy to see that 
\be
H^2 > 0 \quad \Longrightarrow \quad G_+(H^2) > 2\langle M^3/l \rangle>0.
\ee
This means that (\ref{latesym})  has no solutions in $H^2 > 0
$. We conclude that it is impossible to get late time de Sitter
expansion when $\theta_1=\theta_2$.  

Now consider the ``antisymmetric'' equation of motion
\be
0=G_-(H^2) \equiv \Delta F(H^2).
\ee
This is easily solved to give
\be
H^2 \sim H_0^2=-\frac{\Delta(M^6/l^2)}{\Delta{M^6}}
\ee
Therefore,
when $\theta_1=-\theta_2$, we have late time de Sitter expansion
provided $\Delta(M^6/l^2)<0$.

We are ready to ask whether or not this de Sitter phase is preceded by
a period of matter/radiation domination, with $H^2 \propto
\rho$. Consider the ``antisymmetric'' equation of motion (\ref{G-}) at
smaller values of $a$. We can manipulate this equation to give a
quadratic in $H^2+\frac{\ka}{a^2}$,
\begin{multline}
\left(\Delta
M^6\right)^2\left(H^2+\frac{\ka}{a^2}\right)^2+\left[2\Delta
\left[M^6 V(a)\right]\Delta M^6-\frac{\rho^2}{9} \langle M^6 \rangle
\right]\left(H^2+\frac{\ka}{a^2}\right) \\
\left[\Delta \left(M^6V(a)\right)\right]^2+\frac{\rho^2}{36} \left[ \frac{\rho^2}{36}-4 \Big\langle
M^6 V(a) \Big\rangle \right]=0
\end{multline}
where
\be
V(a)=\frac{1}{l^2}-\frac{\mu}{a^4}
\ee
Now solve this quadratic to derive the Friedmann equation
\be
H^2+\frac{\ka}{a^2}=-\frac{\Delta
\left[M^6 V(a)\right]}{\Delta M^6}+\rho^2 \frac{ \langle M^6 \rangle}{18\left(\Delta
M^6\right)^2} \pm\frac{M_1^3M_2^3 \rho}{3 \left(\Delta
M^6\right)^2}\sqrt{\frac{\rho^2}{36}-\Delta M^6 \Delta V}
\ee
The choice of root corresponds to a choice of $\theta= \pm 1$. When
\be
|\Delta \mu/a^4| \ll |\Delta(1/l^2)|  \qquad \textrm{and} \qquad \rho \ll
\rho_\textrm{max}=\frac{6M_1^3M_2^3}{\langle M^6 \rangle}\sqrt{-\Delta(M^6)
\Delta(1/l^2)}
\ee 
The Friedmann equation approximates to
\be \label{FRWapprox}
H^2+\frac{\ka}{a^2}\approx H_0^2 +\frac{\Delta(M^6 \mu)}{\Delta M^6}
\frac{1}{a^4} \pm\frac{M_1^3M_2^3 \sqrt{-\Delta(M^6)
\Delta(1/l^2)}}{3 \left(\Delta
M^6\right)^2} \rho
\ee
Recall that for $H_0^2 >0$, we chose $\Delta M>0$, and demanded that
$\Delta \left(M^6/l^2\right)<0$. This ensures that the square root in equation
(\ref{FRWapprox})  is real. In order to reproduce the Friedmann
equation of the standard cosmology, we must take the positive square
root in (\ref{FRWapprox}), which corresponds to $\theta=1$. We then find that
\be \label{FRWapprox2}
H^2+\frac{\ka}{a^2}\approx H_0^2+\frac{\Delta(M^6 \mu)}{\Delta M^6}
\frac{1}{a^4}+\frac{\rho}{6 M_\textrm{b}^2}
\ee
where the four-dimensional Planck mass on the brane is given by
\be
M_\textrm{b}^2=\frac{\left(\Delta
M^6\right)^2}{2M_1^3M_2^3 \sqrt{-\Delta(M^6)
\Delta(1/l^2)}}
\ee
Note that
\be
\rho_\textrm{max}=\frac{3\left(\Delta
M^6\right)^2}{\langle M^6 \rangle M_\textrm{b}^2}
\ee 
The $\mu/a^4$ term in (\ref{FRWapprox}) comes from the bulk, and
behaves like a form of dark radiation. We can interpret it
holographically as the energy density of a  conformal field theory dual
to the bulk~\cite{Padilla:thesis, Padilla:CFT, Savonije:CFT, Padilla:exact}. For simplicity let us assume that this
contribution is always small compared to the energy density
on the brane, so that the Friedmann equation behaves as
\be \label{FRWapprox3}
H^2+\frac{\ka}{a^2}\approx H_0^2+\frac{\rho}{6 M_\textrm{b}^2}
\ee
When does this equation describe real physics? To predict the
current cosmic acceleration, we need $H_0^2 \sim   10^{-68}
~(\textrm{eV})^2$. Prior to this, we need $H^2 \sim \rho/6m_{pl}^2$,
where $m_{pl} \sim 10^{19}
~\textrm{GeV}$. This must be the case as far back as nucleosynthesis,
at which point $\rho=\rho_\textrm{NS} \sim 10^{24}
~(\textrm{eV})^4$. For equation (\ref{FRWapprox3}) to be physical, we
therefore require that $M_\textrm{b}  \sim m_{pl}$, and
\be \label{cond4}
\frac{\rho_\La}{\rho_\textrm{max}} \ll
\frac{\rho_{\La}}{\rho_\textrm{NS}} \sim 10^{-36}
\ee
where $\rho_\La =6  M_\textrm{b}^2H_0^2 \sim 10^{-12}
~(\textrm{eV})^4$. Note that the scale of curvature in the spatial sections is observed to be very close to zero. This means that we can ignore the $\ka/a^2$ contribution in equation (\ref{FRWapprox3}).

Suppose we take 
\be
M_1=(1+\la)^{1/6}M_2, \qquad l_1=(1+\la+\ep)^{1/2}l_2
\ee
where $\la>0$ is of order one, and $0< \ep \lesssim 10^{-36}$. We find
that
\be
M_\textrm{b}^2 \approx \frac{\la}{2}M_2^3l_2, \qquad H_0^2 \approx
\frac{\ep}{\la(1+\la)} \frac{1}{l_2^2}
\ee
and the condition (\ref{cond4}) holds. As an example, consider $\ep
\sim 10^{-36}$. If we
take $1/l_2 \sim 10^{-16}$ eV and $M_2 \sim 10$ TeV, we
obtain precisely the desired cosmology from nucelosynthesis onwards.

\section{The stringy model} \label{stringy}

Our next example is motivated by the slope expansion in heterotic
string theory. There is no bare cosmological constant in this
expansion and the slope parameter ($\alpha'$) is positive. We therefore take
\be
\La_i=0, \qquad \al_i>0
\ee
As before, we would expect $\mu_i$ to enter the dynamics as some form
of dark radiation (see, for example~\cite{Padilla:gbhol}). For the (+)
branch with $\mu_i>0$,  we must introduce a second brane to shield the
singularity. If this is done at small enough $a$, we would not expect
it to significantly affect the dynamics of the main cosmological
brane. In any case, let us avoid such complications by setting
$\mu_i=0$. We will do this even for the (-) branch to keep our
analysis tidy. 

For the (+) branch, we find  that the bulk metric is given by
\be \label{+}
h(a)=\ka +\frac{a^2}{2 \al} 
\ee
For the (-) branch, we are only allowed $\ka=1$, so that
\be
h(a)=1
\ee
Note that the (+) branch is not well defined at $\al=0$. For this
reason, it represents a significant departure from Einstein gravity,
and is of particular interest. We will focus on this solution
presently.
\subsection{The (+) branch}
For the (+) branch, note that the metric corresponds to anti-de Sitter
space with the appropriate slicing (depending on $\ka$). The effective cosmological constant
is given by
\be
\La_\textrm{eff}=-\frac{3}{\al}.
\ee
This may be surprising given that we set the bare cosmological
constant to zero.   The brane equations of motion are, of course,
given by (\ref{G+}) and (\ref{G-}), but with
\be
F(H^2)=\frac{1}{3}M^3\sqrt{\frac{1}{2 \al}+H^2+\frac{\ka}{a^2}}
\left[1+8 \al \left(H^2+\frac{\ka}{a^2} \right) \right]
\ee
Again, we begin by looking for late time de Sitter solutions. As $a
\to \infty$, 
\be
F(H^2) \to  \frac{1}{3}M^3\sqrt{\frac{1}{2 \al}+H^2}
\left[1+8 \al H^2\right], \qquad \rho \to 0
\ee
The ``symmetric'' equation of motion (\ref{G+}) has no solution, since
\be
H^2 > 0, \quad \al>0 ~\Longrightarrow G_+(H^2) >
\sqrt{2}\langle M^3/\sqrt{\al} \rangle >0
\ee
As before, we conclude that late time de Sitter expansion is
impossible when $\theta_1=\theta_2$.  

Now consider the ``antisymmetric'' equation of motion
(\ref{G-}). Since $G_+(H^2)$ is never zero in $H^2 > 0$, the
equation
\be \label{P}
0=P(H^2) \equiv G_+(H^2)G_-(H^2)=\Delta \left[F(H^2)\right]^2
\ee
must have the same roots ($H_0^2 > 0$) as
$G_-(H^2)=0$. $P(H^2)$ is a cubic in $H^2$,
\be
P(H^2)=\frac{64}{9}\Delta (M^6 \al^2)H^6+\frac{16}{3} \Delta(M^6
\al)H^4+\Delta(M^6)H^2+\frac{1}{18}\Delta\left(\frac{M^6}{\al} \right)
\ee
To ensure the existence of a real solution,
$P(H_0^2)=0$, we demand that
\be \label{cond1}
\Delta\left(\frac{M^6}{\al} \right)<0, \qquad  \Delta(M^6
\al)>0
\ee
This is a sufficient (although perhaps not a necessary) condition. To
see this note that we must have $\al_1>\al_2$ for the two inequalities
in (\ref{cond1}) to be consistent. In addition, we deduce that
$\Delta (M^6 \al^2)>0$. We now see that $P(0)<0$, whereas $P(H^2) \to +\infty$ as $H^2 \to
+\infty$. By the intermediate value theorem, there exists $0< H_0^2<
\infty$, such that $P(H_0^2)=0$.  For $\theta_1=-\theta_2$, we
conclude that we have late time de Sitter expansion whenever the
condition (\ref{cond1}) holds.

We now examine the behaviour of the ``antisymmetric'' equation of
motion (\ref{G-}) at smaller values of $a$. This will provide us with
an estimate for $H_0^2$, and enable us to check for a period of
matter/radiation domination with $H^2 \propto \rho$.

We can manipulate equation  (\ref{G-}) to obtain a polynomial in
$H^2+\frac{\ka}{a^2}$ of degree six. It is therefore very difficult to
find solutions so we adopt another approach. We will expand $G_-(H^2)$
as a Taylor series, about the point $H^2+\frac{\ka}{a^2}=0$,
\be \label{exp}
\theta\frac{\rho}{6}=\frac{1}{3\sqrt{2}}\Delta\left(\frac{~M^3}{\sqrt{\al}}\right)+\frac{~3}{\sqrt{2}}\Delta\left({M^3}{\sqrt{\al}}\right)\left(H^2+\frac{\ka}{a^2}
\right)+\textrm{higher order terms}
\ee
This expansion is valid when
\be \label{expansion}
H^2+\frac{\ka}{a^2} \ll \textrm{min} \Bigg \{ \frac{1}{\al_1},~
\frac{1}{\al_2}, ~\Big | \frac{
\Delta\left({M^3}{\sqrt{\al}}\right)
}{\Delta\left({M^3}{\al\sqrt{\al}}\right)} \Big | \Bigg \}
\ee
Let us assume, for the time being, that this holds. We will come back
to it later. Ignoring the higher order terms, we rearrange equation
(\ref{exp}) to give
\be \label{GBFRWapprox}
H^2+\frac{\ka}{a^2}\approx H_0^2+\frac{\rho}{6 M_\textrm{b}^2}
\ee
where
\be
H_0^2=-\frac{\Delta \left( M^3/\sqrt{\al} \right)}{9\Delta \left
( M^3\sqrt{\al} \right)}, \qquad M_\textrm{b}^2
=\frac{3\Delta \left
( M^3\sqrt{\al} \right)}{\sqrt{2}}
\ee
If we impose condition (\ref{cond1}), we have $H_0^2>0$. Furthermore,
to ensure $M_\textrm{b}^2>0$, we have chosen $\theta=1$. Since we
must now have $\al_1 >\al_2$, it is fairly easy to show that
\be
 \Big | \frac{
\Delta\left({M^3}{\sqrt{\al}}\right)
}{\Delta\left({M^3}{\al\sqrt{\al}}\right)} \Big | <\frac{1}{\al_1} <
\frac{1}{\al_2} 
\ee
Our validity condition now reads
\be \label{valid}
H^2+\frac{\ka}{a^2} \ll \Big | \frac{
\Delta\left({M^3}{\sqrt{\al}}\right)
}{\Delta\left({M^3}{\al\sqrt{\al}}\right)} \Big | 
\ee
For $\rho \gg \rho_\La =6M_\textrm{b}^2H_0^2$, this translates into an
upper bound on the energy density.
\be
\rho \ll \rho_\textrm{max}=6M_\textrm{b}^2 \Big | \frac{
\Delta\left({M^3}{\sqrt{\al}}\right)
}{\Delta\left({M^3}{\al\sqrt{\al}}\right)} \Big |
\ee
For consistency with observations, we again require that $H_0^2 \sim   10^{-68}
~(\textrm{eV})^2$, $M_\textrm{b}  \sim m_{pl}$, and
\be \label{cond5}
\frac{\rho_\La}{\rho_\textrm{max}} \ll
\frac{\rho_{\La}}{\rho_\textrm{NS}} \sim 10^{-36}
\ee
where we remind the reader that $\rho_\textrm{NS} \sim 10^{24}
~(\textrm{eV})^4$ is the energy density at the time of nucleosynthesis.

Suppose we take 
\be
M_1=(1+\la)^{1/6}M_2, \qquad \al_1=\frac{1+\la}{(1-\ep)^2} \al_2
\ee
where, again,  $\la>0$ is of order one, and $0< \ep \lesssim 10^{-36}$. We find
that
\be
M_\textrm{b}^2 \approx \frac{3\la}{\sqrt{2}}M_2^3\sqrt{\al_2}, \qquad H_0^2 \approx
\frac{\ep}{9\la} \frac{1}{\al_2}
\ee
and the condition (\ref{cond5}) holds. As an example, consider $M_2
\sim m_{pl}$, $\al_2 \sim 1/m_{pl}^2$, as we might expect from string
theory. If we take $\ep
\sim 10^{-124}$,  we obtain the desired cosmology after
nucleosynthesis, In fact, with this level of fine tuning, we will
still have
good agreement with the standard cosmology at much earlier times.

We end this section with a few comments about quantum
stability, and the absence of a zero mode. In~\cite{Charmousis:scalar}, there are examples of flat
branes, where the bulk metric corresponds to a (+)
branch. Transverse-tracefree perturbations about these solutions
include a normalisable zero mode. This mode turns out to be a ghost in
the effective theory. We
might be worried that this will also happen here, in the limit that
$H_0^2 \to 0$. However, in our
model, note that $\theta_1=-\theta_2=1$. This means that
$\mathcal{M}_1$ corresponds to  $0 \leqslant a< a(\tau)$, and
$\mathcal{M}_2$ corresponds to $a(\tau) < a< \infty$. The volume of
our background is therefore infinite, so there is no normalisable zero
mode. We cannot, therefore, apply the results
of~\cite{Charmousis:scalar}. 

The absence of a zero mode has another important implication: gravity cannot be localised. However,
given the close relationship
between models with infra-red modifications of gravity and cosmic
acceleration (see, for example, ~\cite{Deffayet:cosmo}) we might expect our model to
exhibit {\it quasi}-localisation. This occurs in the DGP model, where a resonance of massive modes leads to four-dimensional brane gravity at intermediate scales.  Will something similar happen here? Quite possibly. Roughly speaking, in the antisymmetric case, we have finite volume on one side of the brane, and infinite volume on the other. On the finite side we have a localised graviton, whereas on the infinite side we have a non-localised graviton.  At intermediate scales it could be that the finite side dominates, so that gravity appears localised on the brane. This would be consistent with the fact that we reproduce the standard cosmology as far back as nucleosynthesis. Of course, a more thorough investigation is clearly required.

\subsection{The (-) branch}
In this section, we consider the (-) branch, for which $\ka=1, ~h(a)=1$. This
corresponds to flat space in the bulk. The equations of motion are
given by (\ref{G+}) and (\ref{G-}) with
\be
F(H^2)=M^3 \sqrt{\frac{1}{a^2}+H^2} \left[ 1+\frac{8}{3}\al
\left(\frac{1}{a^2}+H^2 \right) \right]
\ee
Now look for late de Sitter behaviour. As $a \to \infty$
\be
F(H^2) \to M^3 H \left[ 1+\frac{8}{3}\al
H^2 \right], \qquad \rho \to 0
\ee
It is easy to see that $G_+(H^2)=0$ has no solution in $H^2>0$. This
rules out the possibility of late time de Sitter when
$\theta_1=\theta_2$. Now consider $G_-(H^2)=0$. This has the solution
\be
H^2=H_0^2=- \frac{3 \Delta M^3}{8 \Delta (M^3 \al)}
\ee
This can be made positive if we take
\be
\Delta M^3 >0, \qquad \Delta (M^3 \al)<0
\ee
At smaller values of $a$, equation (\ref{G-}) is already a cubic in
$\mathcal{H}= \sqrt{\frac{1}{a^2}+H^2}$,
\be
\frac{8}{3} \Delta (M^3 \al) \mathcal{H}^3+\Delta M^3
\mathcal{H}-\theta \frac{\rho}{6}=0
\ee
For $\theta=1$, this equation has no roots with $\mathcal{H} \geqslant
H_0$. We therefore choose $\theta=-1$, and find that the only real root
with  $\mathcal{H} \geqslant
H_0$ is given by
\be \label{cuberoot}
\mathcal{H}=H_++H_-
\ee
where 
\be
H_{\pm}^3= H_\rho^3 \pm \sqrt{H_\rho^6-\frac{H_0^6}{27}}, \qquad
-\frac{\pi}{3} \leqslant \textrm{arg}\left(H_{\pm}\right) \leqslant \frac{\pi}{3}
\ee 
and 
\be
H_\rho^3=-\frac{\rho}{32  \Delta (M^3 \al)}.
\ee
During a matter/radiation dominated era, we expect $H_\rho \gg H_0$,
so that
\be
\mathcal{H} \approx 2^{1/3}H_\rho
\ee
This implies that 
\be
H^2+\frac{1}{a^2} \propto \rho^{2/3}
\ee
which disagrees with the standard cosmology. We conclude that this
model does give cosmic acceleration, but does not predict the
correct physics beforehand. Perhaps this is not surprising given that we
have a flat bulk. For gravity to be localised on
the brane we would expect there to be a warp factor in the bulk metric.

\section{Discussion} \label{Discussion}
In this paper, we have considered the cosmology of an asymmetric
3-brane, sandwiched between two different bulk spacetimes. In general,
the bulk spacetimes are solutions to Gauss-Bonnet gravity, with the
gravitational couplings being allowed to differ on either side of the
brane. This leads to weighted junction conditions across the brane, so
that we naturally have many more solutions available.

We focussed on two special cases: a generalized Randall-Sundrum model
and a stringy model. In the former, we switched off all higher
derivative couplings, and set the bare cosmological constant to be
negative. In the latter, we set the bare cosmological constant to be
zero, and demanded that the Gauss-Bonnet couplings be positive, in
accordance with string theory. 

The generic behaviour of both models
was the same, depending crucially on whether we were considering a
``symmetric'' or an ``antisymmetric'' scenario. For the symmetric scenario, $a$ was a minimum (or a
maximum) at
the brane for both bulk spacetimes. For the antisymmetric scenario,
$a$ was a maximum at the brane for one side of the bulk, and a minimum
for the other. The symmetric scenario did not permit de
Sitter cosmologies, unlike the antisymmetric scenario. Naturally, to ensure that the late time acceleration
agreed with observations, a degree of fine-tuning was required. What
is interesting is that cosmic acceleration of any sort could be
achieved without resorting to vacuum energy or induced curvature. It
was made possible by the weighted junction conditions. Furthermore,
for the antisymmetric case, the brane cosmology could be made to follow the standard
cosmology all the way back to nucleosynthesis. This is our main
result. 

Finally, as promised, we will comment on the initial singularity. As
they are, our models will predict a time of infinite energy density,
so that our theory will eventually break down. We might hope that by adjusting the
bulk mass parameters,  we could enforce a ``bounce'' cosmology. This means
that the scale factor has a non-zero minimum. However, the bulk mass
looks like dark radiation on the brane, so it is difficult to see what
new effect it could have. However, in Einstein gravity, bounce cosmologies {\it do} occur
for branes moving in between {\it charged} black holes~\cite{Padilla:exact,
Padilla:thesis}. We could consider the analagous situation in
Gauss-Bonnet gravity coupled to electromagnetism. Can the initial
singularity be avoided without spoiling the late time behaviour? We
will leave this for future research.

\vskip .5in

\centerline{\bf Acknowledgements}
\medskip
I would like to thank Christos Charmousis, Syksy Rasanen, John March-Russell, and especially James Gregory for useful discussions.  AP was funded by PPARC

\bibliographystyle{utphys}

\bibliography{gbacc}

\end{document}